\documentclass{cimento}

\usepackage{graphicx}  % got figures? uncomment this

\title{$K^+\rightarrow\pi^+\nu\bar \nu$ decay and NP searches at NA62}
\author{S.~Martellotti\from{ins:x} on behalf of the NA62 Collaboration}
\instlist{\inst{ins:x} INFN, Laboratori Nazionali di Frascati - Frascati, Italy}
\begin{document}

\maketitle

\begin{abstract}
$K\rightarrow \pi\nu \bar\nu$ is one of the theoretically cleanest meson decay where to look for indirect effects of new physics complementary to LHC searches. The NA62 experiment at CERN SPS is designed to measure the branching ratio of the $K^+\rightarrow \pi^+\nu \bar\nu$ decay with 10\% precision. 
After commissioning, in 2016 first data set good for physics has been collected. The data taking is foreseen till the LS2 at the end of 2018.
The analysis method has been tested with 5\% of the full 2016 statistics. Results obtained with this subsample will be shown.
The status of the experiment will be reviewed, and prospects will be presented, together with a mention to the broader NA62 physics program.

PACS - 13.20Eb Decays of K mesons.
\end{abstract}

\section{Introduction}

\begin{figure}[h]
\begin{center}
\includegraphics[width=10cm]{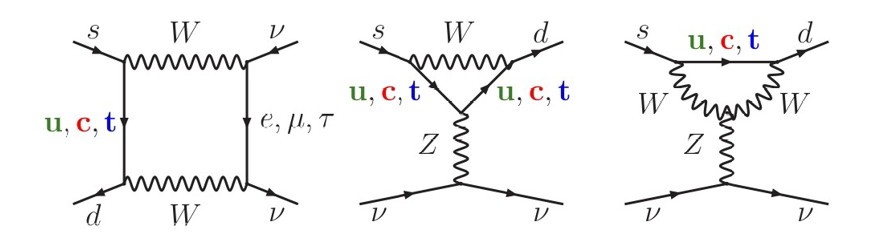}
\caption{Box diagram and Z-penguin diagrams contributing to the process $K\rightarrow \pi\nu\bar\nu$.}\label{fig:1}
\end{center}
\end{figure}

The $K^+\rightarrow \pi^+ \nu \bar \nu$ and $K_L\rightarrow \pi^0 \nu \bar \nu$ are 
flavour changing neutral current decays proceeding through box and electroweak penguin diagrams as shown in fig.\ref{fig:1}. A quadratic GIM mechanism and strong Cabibbo suppression make these processes extremely rare. Using the value of tree-level elements of the Cabibbo-Kobayashi-Maskawa (CKM) triangle as external inputs, the Standard Model (SM) predicts\cite{ref:carico}\cite{ref:neutro}:
\begin{displaymath}
BR(K^+\rightarrow \pi^+ \nu \bar \nu) = (8.4 \pm 1.0) \times 10^{-11}, \,
BR(K_L\rightarrow \pi^0 \nu \bar \nu) = (3.4 \pm 0.6) \times 10^{-11}.
\end{displaymath}

The theoretical accuracy is at the percent level, because short distance physics
dominates thanks to the top quark exchange in the loop. The hadronic matrix elements cancel
almost completely in the normalization of the $K \rightarrow \pi\nu\bar\nu$
branching ratios to the precisely measured BR($K^+ \rightarrow \pi^0 e^+ \nu_e$).
Experimental knowledge of the external inputs dominate the uncertainties on these predictions. 
The dependence on CKM parameters partially cancels in the correlation between 
$K^+\rightarrow \pi^+ \nu \bar \nu$ and $K_L\rightarrow \pi^0 \nu \bar \nu$.
Therefore simultaneous measurements of the two BRs would allow a theoretically clean investigation of the CKM triangle using kaons only. The $K \rightarrow \pi\nu\bar\nu$ decays are
extremely sensitive to physics beyond the SM, probing the highest mass scales among the rare 
meson decays. The largest deviations from SM are expected in
models with new sources of flavour violation, owing to weaker constraints from B physics 
\cite{ref:teoria1}\cite{ref:teoria2}.
The experimental value of $\epsilon_K$,
the parameter measuring the indirect CP violation in neutral kaon decays, limits the range of variation expected for $K \rightarrow \pi\nu\bar\nu$ BRs within 
models with currents of defined chirality, producing typical correlation patterns between charged and neutral
modes\cite{ref:teoria3}. 
Results from LHC direct searches strongly limit the range of variation, mainly in supersymmetric models\cite{ref:ss1}\cite{ref:ss2}. 
In any case, due to the suppression of this decay in the SM, significant variations of the $K \rightarrow \pi\nu\bar\nu$ BRs from the SM predictions
induced by new physics at mass scales up to 100 TeV are still observable by experiment with at least 10\% precision, even with existing constraints from other measurements in K physics. 

The most precise experimental result has been obtained by the dedicated experiments E787 and E949 at the Brookhaven National Laboratory which collected a total of 7 events using a decay-at-rest technique. Only the charged mode has been observed so far, and the present experimental status is\cite{ref:misura1}\cite{ref:misura2}\cite{ref:misura3}:
\begin{math}
BR(K^+\rightarrow \pi^+ \nu\bar \nu)_{exp} = (17.3^{+11.5}_{-10.5}) \times 10^{-11}, \,
BR(K_L\rightarrow \pi^0 \nu \bar \nu) _{exp} < 2.6 \times 10^{-8}  \; 90\% CL,
\end{math}
still far from the precision of the SM prediction.

\section{The NA62 experiment at CERN.}

The NA62 experiment at CERN\cite{ref:exp1}\cite{ref:exp2} aims to measure the $BR(K^+\rightarrow \pi^+ \nu \bar \nu)$ with 10\% precision. Therefore it needs to collect about $10^{13}$ $K^+$ decays using 400 GeV/c protons from SPS for a 10\% signal acceptance. Keeping the background to signal ratio about 10\% requires the use of
almost independent experimental techniques to suppress unwanted final states. With a single event
sensitivity of $10^{-12}$ NA62 can afford also a broader physics program\cite{ref:rug}. NA62 is running with
the apparatus fully operational since 2016.

\begin{figure}[h]
\begin{center}
\includegraphics[width=14cm]{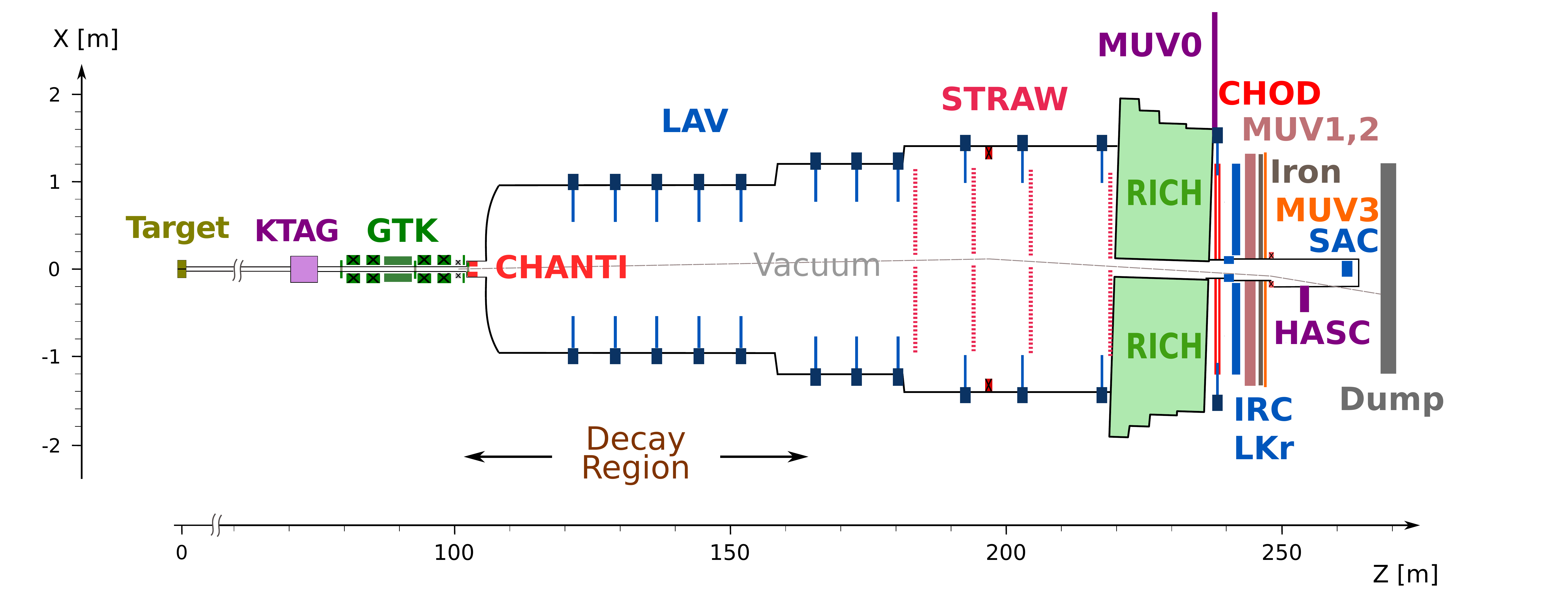}
\caption{\label{fig:setup}Schematic layout of the NA62 experiment in the xz plane.}
\end{center}
\end{figure}

NA62 adopts a kaon decay-in-flight technique. Fig.\ref{fig:setup} shows a schematic view of the apparatus.
Primary SPS protons strike a target from which a secondary charged hadron beam of 75 GeV/c and
1\% momentum bite is selected and transported to the decay region. The detailed descriptions of
the apparatus can be found in\cite{ref:exp_new}. The incoming kaon is positively identified by a differential
Cerenkov counter (KTAG) and its momentum and direction are measured by three stations of Si
pixel detectors (GTK). About 6\% of beam particles are $K^+$. A guard ring detector (CHANTI)
vetoes beam inelastic interactions occurring in GTK. A decay tank at vacuum ($10^{-6}$ mbar)
is surrounded by ring-shaped lead-glass calorimeters designed to intercept photons at polar angles of up to 50 mrad (LAV). Four stations of straw chambers (STRAW) in vacuum track downstream charged particles, with a dipole magnet providing a 270 MeV/c transverse kick for momentum analysis. A RICH counter time-stamps and identifies charged particles; plastic scintillators (CHOD) are used for triggering
and timing. Photon rejection in the forward region is provided by an electromagnetic calorimeter
of liquid krypton (LKr) and two small angle calorimeters (IRC and SAC). Hadron calorimeters (MUV1,2)
and a plastic scintillator detector (MUV3) are used to suppress muons. At full intensity, the SPS delivers $3.3 \times 10^{12}$ protons per pulse to NA62, corresponding to a particle rate of 750 MHz in the GTK.
Information from CHOD, RICH, MUV3 and LKr are built up online to issue level zero trigger
conditions. Software-based variables from KTAG, CHOD, LAV and STRAW provide higher level
trigger requirements. $K^+\rightarrow \pi^+ \nu \bar \nu$-triggered data are taken concurrently with downscaled samples of data for rare kaon decays studies and minimum bias.
The NA62 apparatus has been commissioned in 2015 and 2016. Low intensity data have been
taken in 2015 with a minimum bias trigger to study detector performances and to perform physics
analysis. In fall 2016 NA62 has collected about $4.5 \times 10^{11}$ kaon decays for $K^+\rightarrow \pi^+ \nu \bar \nu$ at 20-40\% of nominal intensity. A four-month run dedicated to $K^+\rightarrow \pi^+ \nu \bar \nu$ has taken place in 2017 at 50-60\% of the nominal intensity and the experiment is currently running for the last 2018 run, for which seven months of data taking at 60\% of the nominal intensity are scheduled.

\section{ The $K^+\rightarrow \pi^+ \nu \bar \nu$ analysis.} 
The analysis of 5\% of the 2016 dataset corresponding to $2.3 \times 10^{10}$ kaons is presented here.
The $K^+\rightarrow \pi^+ \nu \bar \nu$ signature is one track in the initial and final state with two missing neutrinos. The main kinematic variable is $m^2_{miss} = (P_{K} - P_{\pi})^2$, where $P_K$ and $P_\pi$ are the 4-momenta of the $K^+$ and $\pi^+$ respectively.

\begin{figure}[h]
\begin{center}
\includegraphics[width=17pc]{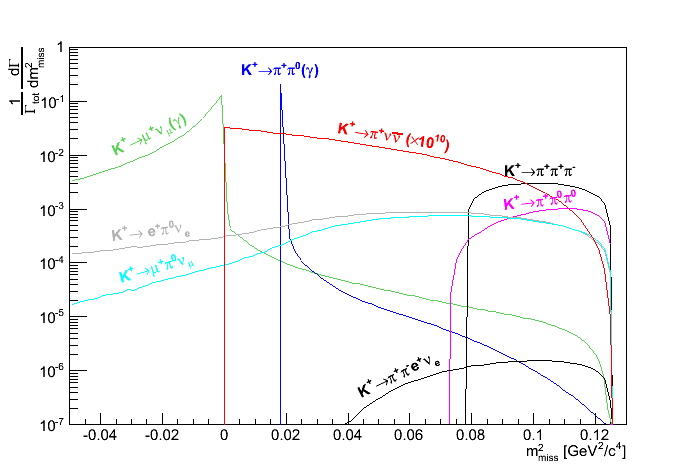}
\caption{\label{fig:missLog} $m^2_{miss}$ distributions for signal and backgrounds of the main $K^+$ decay modes are shown in log scale: the backgrounds are normalized according to their branching ratio; the signal is multiplied by a factor $10^{10}$.}
\end{center}
\end{figure} 

The theoretical shapes of the $m^2_{miss}$ distribution for the main $K^+$ background decay modes
are compared to the $K^+\rightarrow \pi^+ \nu \bar \nu$ on fig.\ref{fig:missLog}. The analysis is done in the $\pi^+$ momentum range between 15 and 35 GeV/c to leave at least 40 GeV of electromagnetic energy in the calorimeters in the case of $K^+\rightarrow \pi^+ \pi^0 (K_{\pi2})$ decay. Two regions are used: region 1 between $K^+\rightarrow \mu^+ \nu_\mu (K_{\mu2})$ and $K_{\pi2}$ and region 2 between $K_{\pi2}$ and $K^+\rightarrow \pi^+ \pi^+ \pi^- (K_{\pi3})$. The main backgrounds entering those regions are $K_{\mu2}$ and $K_{\pi2}$ decays through non gaussian resolution and radiative tails; $K_{\pi3}$ through non-gaussian resolution; $K^+\rightarrow \pi^+ \pi^- e^+ \nu_e (K_{e4})$ and $K^+\rightarrow \l^+ \pi^0 \nu_l (K_{l3})$ by not detecting the extra $\pi^-, e^+, \pi^0$ particles. Another important source of background is the beam-related background coming from upstream decays and beam-detector interactions. Each of the background processes requires a different rejection procedure depending on its kinematics and type of charged particle in the final state. The main requirements for the analysis are excellent kinematic reconstruction to reduce kinematic tails; precise timing to reduce the kaon mis-tagging probability; no extra in-time activity in all of the electromagnetic calorimeters to suppress $K^+\rightarrow \pi^+\pi^0$ decays with $\pi^0 \rightarrow \gamma\gamma$ (photon rejection); clear separation between $\pi/\mu/e$ tracks to suppress decays with $\mu^+$ or $e^+$ in the final state (particle identification). Low multiplicity cuts in the downstream detectors are used to further suppress decays with multiple charged tracks in the final state.

\begin{figure}[h]
\begin{center}
\includegraphics[width=7cm]{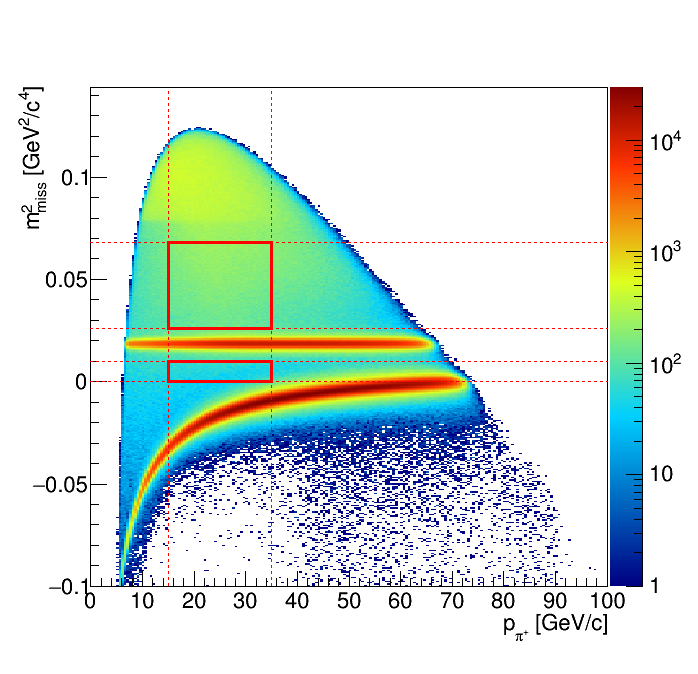}
\caption{Distribution of $m^2_{miss}$ as a function of pion momentum for kaon events selected on control data. The signal regions (red box) in the ($m^2_{miss}$, $P_{\pi^+}$) plane are drawn for reference.}\label{fig:missData}
\end{center}
\end{figure}

The parent $K^+$ track is reconstructed and time-stamped in the GTK with 100 ps resolution; the daughter
$\pi^+$ track is reconstructed in the STRAW. The CHOD and RICH measure $\pi^+$ time with resolution below 100 ps. The pion is associated in time to a KTAG kaon signal. The timing and the closest distance
of approach between GTK and STRAW tracks allow a precise $K^+ - \pi^+$ matching. The kaon mistagging
probability at 40\% of nominal intensity is below 2\%, signal acceptance about 75\%.

\begin{figure}[htbp]
\begin{minipage}{0.5\textwidth} 
\includegraphics[width=6.5cm]{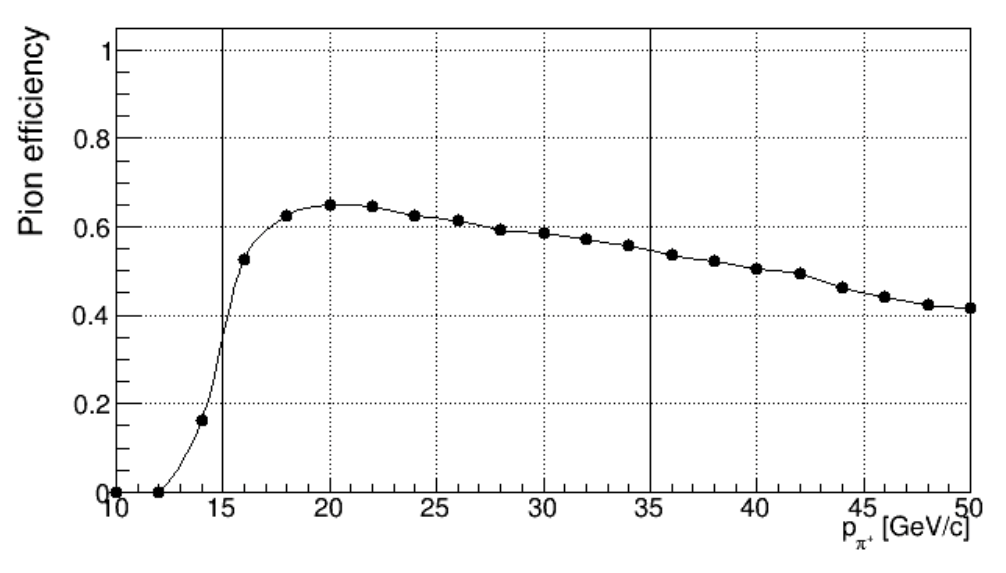} 
\end{minipage} 
\hfil
\begin{minipage}{0.5\textwidth} 
\includegraphics[width=6.5cm]{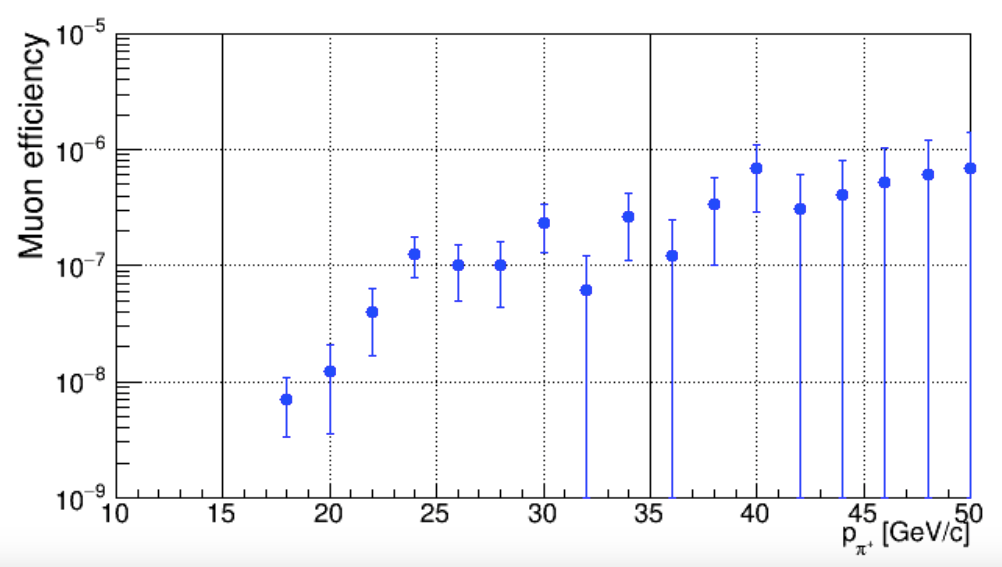}
\end{minipage} 
\caption{Combined pion (left) and muon efficiency (right) after pion identification in calorimeter and RICH, assuming that the two detectors are independent.} 
\label{fig:pion_id}
\end{figure}

Decays are selected within a 50 m fiducial region beginning 10 m downstream of the last
GTK station (GTK3) to reject events originated from interactions of beam particles in GTK and
kaon decays upstream of GTK3. Fig. \ref{fig:missData} (left) exemplifies the kinematics of the selected events. The resolution of $m^2_{miss}$ drives the choice of the boundaries of the signal regions. Reconstruction tails from $K^+\rightarrow \pi^+ \pi^0$, $K^+\rightarrow \mu^+ \nu_\mu$ and $K^+\rightarrow \pi^+ \pi^+ \pi^-$ set the level of background in signal
regions. To reduce it, signal regions are restricted to boxes within a 3D space, defined by i) $m^2_{miss}$; ii)
the same quantity computed using the momentum of the particle measured by the RICH under
$\pi^+$ hypothesis, rather than the straws ($m^2_{miss}$(RICH)); iii) the same quantity computed replacing the 3-momentum of the kaon measured by the GTK with the nominal 3-momentum of the beam $m^2_{miss}$(No-GTK). 
The probability for
$K^+\rightarrow \pi^+ \pi^0$ ($K^+\rightarrow \mu^+ \nu_\mu$) to enter the signal regions is $6 \times 10^{-4}$ ($3 \times 10^{-4}$), as measured with data.
Calorimeters and RICH separate $\pi^+, \mu^+$, and $e^+$. A multivariate analysis combines calorimetric
information and provides $10^5$ $\mu^+$ suppression and 80\% $\pi^+$ efficiency. RICH quantities are used to infer particle types, giving $10^2$ $\mu^+$ suppression and 80\% $\pi^+$ efficiency. The two methods are independent and therefore able to suppress $\mu^+$ by 7 orders of magnitude while keeping 65\% of $\pi^+$. The final $\pi^+$, $\mu^+$ separation performances are summarised in fig. \ref{fig:pion_id}.

\begin{figure}[h]
\begin{center}
\includegraphics[width=7cm]{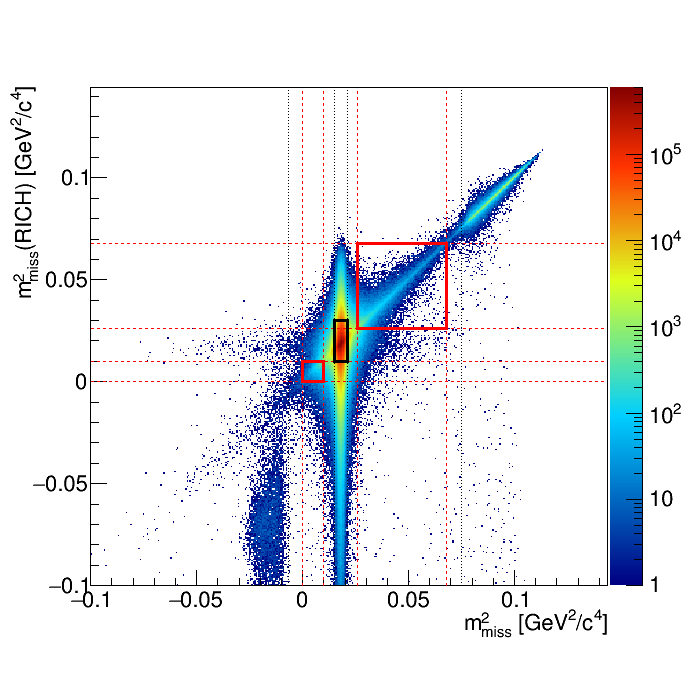}
\caption{Distribution in the ($m^2_{miss}$(RICH), $m^2_{miss}$) plane of events triggered by the PNN trigger line and passing the $\pi\nu\nu$ selection before the photon rejection and with 15 < $P_{\pi^+}$ < 35 GeV/c. The signal regions (red tick box) and the $\pi\pi$ region (black tick box) are drawn.}\label{fig:diffMiss}
\end{center}
\end{figure}

Remaining events after $\pi^+$ identification are primarily $K^+\rightarrow \pi^+ \pi^0$. Fig. \ref{fig:diffMiss} shows the $m^2_{miss}$ distribution of the PNN triggered events surviving the selection before the photon rejection in the ($m^2_{miss}$(RICH), $m^2_{miss}$) plane.

\begin{figure}[htbp]
\begin{minipage}{0.5\textwidth} 
\includegraphics[width=6cm]{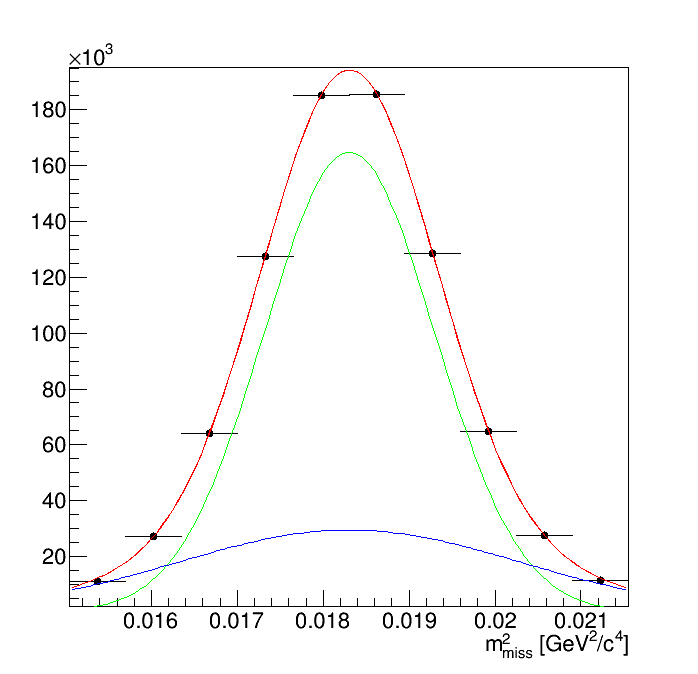} 
\end{minipage} 
\hfil
\begin{minipage}{0.5\textwidth} 
\includegraphics[width=6cm]{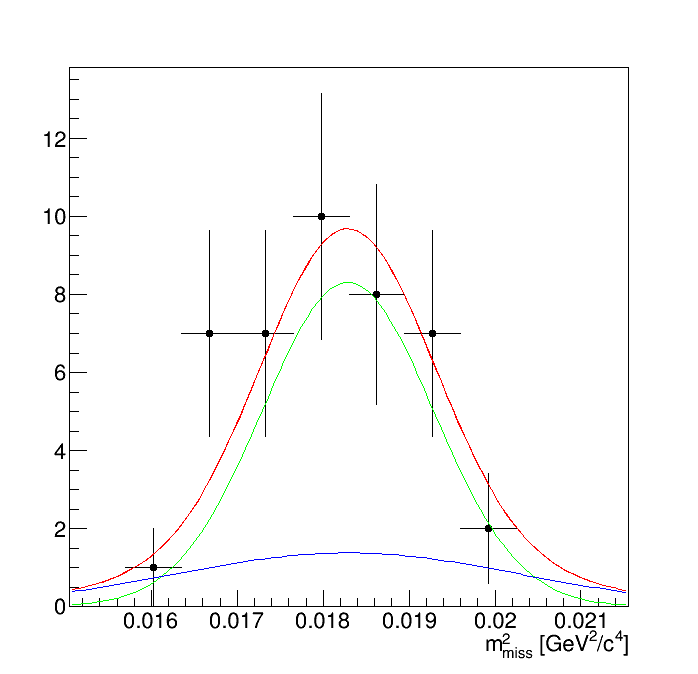}
\end{minipage} 
\caption{Distribution of events in the $\pi\pi$ region before (left) and after (right) photon rejection. Events before photon rejection are from control trigger and must be multiplied by the downscaling factor (400) to get the total number of $K^+\rightarrow \pi^+\pi^0$ before photon rejection. The 2-gaussian fit is superimposed. The overall scale is the only free parameter in the fit of the events after photon rejection, the other parameters are fixed to the value obtained by the fit before photon rejection.} 
\label{fig:rejection}
\end{figure}
Photon rejection exploits timing coincidences between $\pi^+$ and calorimetric deposits. The $\pi^0$ rejection inefficiency is measured from minimum bias and $K^+\rightarrow \pi^+ \nu \bar \nu$-triggered events before and after $\gamma$ rejection, respectively.
Two gaussian functions fit the $m^2_{miss}$ peak of events in the $\pi^+\pi^0$-region before photon
rejection (fig. \ref{fig:rejection} left). The same functions are used to fit
the $m^2_{miss}$ peak of events in the $\pi^+\pi^0$-region after photon rejection (fig. \ref{fig:rejection}  right); in this case the only free parameter of the fit is the overall normalization, with the means,
standard deviations and relative ratio of the two gaussian functions kept fixed to the values
obtained by the fit before photon rejection. The statistics of this sample translates to about $3.3 \times 10^8$ $K^+ \rightarrow \pi^+\pi^0$ events selected before photon rejection and 41 after, corresponding to a $\pi^0$ rejection efficiency of $(1.2 \pm 0.2) \times 10^{-7}$.
Random losses are in the 15-20\% range.

\begin{figure}[h]
\begin{center}
\includegraphics[width=7cm]{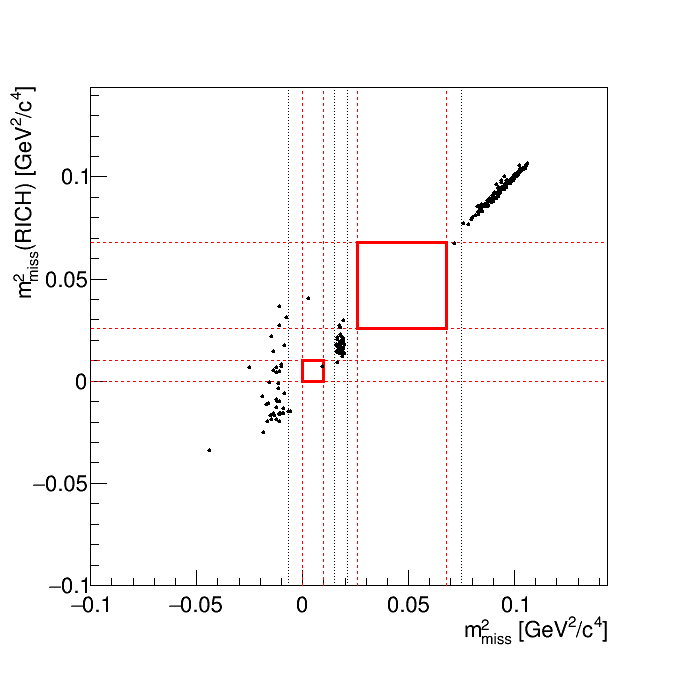}
\caption{Distribution in the ($m^2_{miss}$(RICH), $m^2_{miss}$) plane of $K^+\rightarrow \pi^+ \nu \bar \nu$-triggered events passing the selection, except for the cut on $m^2_{miss}$(No-GTK); signal regions (red tick boxes) and lines defining background regions (light dashed lines) are drawn; the event in region 1 has $m^2_{miss}$(No-GTK) outside the signal region.}\label{fig:final}
\end{center}
\end{figure}

A sample of $K^+\rightarrow \pi^+ \pi^0$ from minimum bias is used for normalization. About 0.064 $K^+\rightarrow \pi^+ \nu \bar \nu$ events are expected over $2.3 \times 10^{10}$ $K^+$ decays. Fig. \ref{fig:final} shows the distribution of residual events in the $m^2_{miss}$(RICH) versus $m^2_{miss}$ plane. 
The event appearing in region 1 has $m^2_{miss}$(No-GTK) outside the signal region. No events are observed in signal regions. Taking into account the signal acceptance, the single event sensitivity turns out to be below $10^{−9}$.
Background estimation from $K^+\rightarrow \pi^+ \pi^0$, $K^+\rightarrow \mu^+ \nu_\mu$ and $K^+\rightarrow \pi^+ \pi^+\pi^-$ are 0.024, 0.011 and 0.017, respectively. They are estimated directly from events outside signal regions, with the measured kinematic tails used for extrapolation in signal regions.
Simulation studies indicate that background from other processes is lower or negligible. 
With this small 2016 data subsample the analysis method has been tested and validated.

\section{NA62 broader physics program}
The high-intensity setup, trigger system flexibility, and detector performance - high-frequency tracking of beam particles, redundant PID, ultra-high-efficiency photon vetoes - make NA62 particularly suitable for searching for new-physics effect from different scenarios.

As for the $K^+\rightarrow \pi^+ \nu \bar \nu$ decay, the NA62 sensitivities to the branching ratios of most rare and forbidden decays of the $K^+$ and $\pi^0$ are in the range $10^{−11}$ to $10^{−12}$ (the main source of neutral pions being $K^+\rightarrow \pi^+\pi^0$ decay). Advanced particle identification, background suppression capabilities and excellent resolution open a way for a programme of rare decay measurements and searches for forbidden decays at record precision. 
The rare decay programme includes measurements of $K^+\rightarrow\pi^+l^+l^−$, $K^+\rightarrow \pi^+\gamma l^+l^−\, (l = e,\mu)$, $K^+ \rightarrow \pi^+\pi^0 e^+e^−$ and $\pi^0 \rightarrow e^+e^−$ processes. The forbidden decays under consideration include lepton flavour and lepton number violating decays (LFV, LNV) like, $K^+\rightarrow\pi^−l_1 l_2$, $K^+\rightarrow \bar l_1\bar\nu l^+_2 l^+_2 (l_{1,2} = e, \mu)$ and $\pi^0 \rightarrow \mu^\pm e^\pm$ decays, as well as $K^+ \rightarrow \pi^+\pi^+l^−\nu$ decays violating the $\Delta S = \Delta Q$ selection rule.

The NA62 beam intensity corresponds to a large number of proton interactions in the T10 target: one nominal year of data taking corresponds to a few $10^{18}$ protons on target (POT), which in turn corresponds to  $10^{18}$  mesons with the following ratios per species, $\pi^0/ \eta/\eta\prime/\Phi/\rho/\omega$ = 6.4/0.68/0.07/0.03/0.94/0.95, and to $\sim10^{15}$ charmed mesons. The mesons, in turn, might decay to exotic particles (dark photons, heavy neutral leptons). These are expected to have feeble interactions with SM fields and therefore to be extremely long lived: an exotic particle produced in the target might reach the NA62 sensitive volume, decaying therein.
Dedicated triggers have been implemented in order to search for two-body leptonic and semi-leptonic decays of dark photons or heavy neutral leptons.
An analysis is also ongoing to search for dark photon in the $\pi^0$ decay via $K^+ \rightarrow \pi^+\pi^0$ with $\pi^0 \rightarrow A^\prime \gamma$, the analysis is synergic with and parasitic to the $K^+\rightarrow \pi^+ \nu \bar \nu$ measurement.

We can also search for Axion-like particles predominantly coupled to two photons and produced in the upstream copper collimator through a Primakov process. In 2016 we dedicated about 17 hours of data taking to this search. In these runs the T10 target was removed and the P42 proton beam was dumped entirely into the collimators.

\section{Conclusions}
The kaon experiment NA62 at CERN is running to search for physics
beyond the SM through the ultra rare $K^+\rightarrow \pi^+ \nu \bar \nu$ decay. The performance of the experiments is within expectations. No events were found after the analysis of $2.3\times 10^{10}$ $K^+$ decays corresponding to 5\% of the full 2016 dataset. 
NA62 is expected to reach the SM sensitivity ($\mathcal{O}(1)$) from the analysis of the 2016 full dataset and to select some tens of $K^+\rightarrow \pi^+ \nu \bar \nu$ events from the analysis of the data taken in 2017 and 2018 runs. The analysis
is on-going together with an optimization of the selection to further reduce backgrounds
and increase signal acceptance. Together with the main measurement, a compelling broader physics program is going to be addressed.

\end{document}